\documentclass[12pt,preprint2,usenatbib,useAMS]{emulateapj}

\shorttitle{}
\shortauthors{Davies et al.}
\usepackage{graphicx}
\usepackage{natbib}
\newcommand{\um}{$\mu$m}

\def\hii{H\,{\sc ii}}

\def\mgi{Mg\,{\sc i}}
\def\afe{[$\alpha$/Fe]}
\def\fei{Fe\,{\sc i}}
\def\sii{Si\,{\sc i}}
\def\tii{Ti\,{\sc i}}
\def\logg{$\log g$}
\def\logz{$\log Z$}
\def\chisq{$\chi^{2}$}
\def\ga{\mathrel{\hbox{\rlap{\hbox{\lower4pt\hbox{$\sim$}}}\hbox{$>$}}}}
\def\la{\mathrel{\hbox{\rlap{\hbox{\lower4pt\hbox{$\sim$}}}\hbox{$<$}}}}

\def\msun{$M$\mbox{$_{\normalsize\odot}$}}

\def\lsun{$L$\mbox{$_{\normalsize\odot}$}}
\def\zsun{$Z$\mbox{$_{\normalsize\odot}$}}
\def\kms{\,km~s$^{-1}$}

\def\arcsec{$^{\prime \prime}$}

\def\teff{$T_{\rm eff}$}

\newcommand{\fig}[1]{Fig.\ \ref{#1}}

\begin{document}
\title{Red Supergiants as cosmic abundance
  probes: \\ the Magellanic Clouds }
\author{Ben Davies\altaffilmark{1,5}, Rolf-Peter
  Kudritzki\altaffilmark{2,3}, Zach Gazak\altaffilmark{2}, 
  Bertrand Plez\altaffilmark{4}, Maria Bergemann\altaffilmark{5,6},\\
  Chris Evans\altaffilmark{7}, Lee Patrick\altaffilmark{8} }

\affil{$^{1}$Astrophysics Research Institute, Liverpool John Moores University, Liverpool Science Park ic2, \\ 146 Brownlow Hill, Liverpool L3 5RF, UK.}


\affil{$^{2}$Institute for Astronomy, University of Hawaii, 2680
Woodlawn Drive, Honolulu, HI, 96822, USA} 

\affil{$^{3}$University Observatory Munich, Scheinerstr. 1, D-81679
Munich, Germany}

\affil{$^{4}$Laboratoire Univers et Particules de Montpellier, Universit\'{e}
Montpellier 2, CNRS, F-34095 Montpellier, France}

\affil{$^{5}$Institute of Astronomy, University of Cambridge, Madingley
Road, Cambridge CB3 0HA, UK}

\affil{$^{6}$Max-Planck Institute for Astronomy, Koenigstuhl 17, Heidelberg, D-69117, Germany}

\affil{$^{7}$ UK Astronomy Technology Centre, Royal Observatory Edinburgh, Blackford Hill, Edinburgh, EH9 3HJ, UK} 

\affil{$^{8}$ Institute for Astronomy, University of Edinburgh, Blackford Hill, Edinburgh, EH9 3HJ, UK}

\begin{abstract}
Red Supergiants (RSGs) are cool ($\sim 4000$K), highly luminous stars ($L \sim 10^{5}$\lsun), and are among the brightest near-infrared (NIR) sources in star-forming galaxies. This makes them powerful probes of the properties of their host galaxies, such as kinematics and chemical abundances. We have developed a technique whereby metallicities of RSGs may be extracted from a narrow spectral window around 1\um\ from only moderate resolution data. The method is therefore extremely efficient, allowing stars at large distances to be studied, and so has tremendous potential for extragalactic abundance work. Here, we present an abundance study of the Large and Small Magellanic Clouds (LMC and SMC respectively) using samples of 9-10 RSGs in each. We find average abundances for the two galaxies of $[Z]_{\rm LMC} = -0.37 \pm 0.14$ and  $[Z]_{\rm SMC} = -0.53 \pm 0.16$ (with respect to a Solar metallicity of \zsun=0.012). These values are consistent with other studies of young stars in these galaxies, and though our result for the SMC may appear high it is consistent with recent studies of hot stars which find 0.5-0.8dex below Solar. Our best-fit temperatures are on the whole consistent with those from fits to the optical-infrared spectral energy distributions, which is remarkable considering the narrow spectral range being studied. Combined with our recent study of RSGs in the Galactic cluster Per~OB1, these results indicate that this technique performs well over a range of metallicities, paving the way for forthcoming studies of more distant galaxies beyond the Local Group.


\end{abstract}



\section{Introduction} \label{sec:intro}
 The observed relationship between a galaxy's stellar mass and its
central metallicity, as well as the abundance trends as a function of
galactocentric distance in spiral galaxies, have provided vital clues
as to how galaxies form and evolve both in the local universe
\citep[e.g.][]{Zaritsky94,Garnett02} and at larger redshift
\citep[e.g.][]{Tremonti04,Erb06,Maiolino08}. These observations have
been used to test the theoretical predictions of various aspects of
galaxy formation and evolution under the framework of a dark energy
and cold dark matter dominated universe, such as hierarchical
clustering, infall, galactic winds and variations in the stellar
Initial Mass Function (IMF) \citep[][plus many
  others]{DeLucia04,Koeppen07,Colavitti08,Dave11}.

However, deriving a galaxy's abundances is a non-trivial task. Most commonly
it has been done by measuring the strengths of \hii-region emission lines. A problem is encountered when metallicities approach the Solar value, in that the temperature-sensitive `auroral' lines become extremely faint, and instead one must rely on empirically calibrated ratios of strong lines. However, many such calibrations exist, and they are known to have large systematic differences in particular at high metallicities. This has a profound effect on the reliability of the mass-metallicity relation and internal abundance gradients  of galaxies, both crucial diagnostics of galaxy evolution \citep[][]{K-E08,Bresolin09}. 

A promising alternative metallicity probe to \hii-regions are evolved massive stars. They are highly luminous ($\sim 10^5$\lsun), allowing them to be studied at Mpc distances. Correct interpretation of their spectra is made possible by sophisticated model atmospheres, and so such work is free of the systematic effects that hamper studies of \hii-regions such as unknown temperature structures and large-scale inhomogeneities. A small sample of nearby galaxies have been studied using blue supergiants \cite[BSGs, e.g.][]{Kudritzki12,Kudritzki13,Kudritzki14}, and it was shown in the case of NGC~300 that the BSG results were in excellent agreement with those of auroral line \hii-region measurements \citep{Bresolin09}. 

Another promising probe of extra-galactic abundances are Red Supergiants (RSGs). These stars are luminous ($L \ga 10^5$\lsun), and with fluxes which peak at $\sim$1\um\ are among the brightest NIR sources in a galaxy. Their brightnesses and colours make them extremely easy to identify, with several thousand such stars expected in a Milky-Way type galaxy. Their young ages ($\la20$Myr) mean that their surface abundances are representative of those in the interstellar medium, aside from those of carbon and nitrogen which are slightly modulated by CNO burning \citep{RSGCabund}. Historically, abundance studies of RSGs have required spectra with high resolving powers ($\lambda/\Delta \lambda \equiv R \ga 20,000$) in order to distinguish the diagnostic metallic lines from the millions of overlapping molecular lines \citep[e.g.][]{Cunha07,RSGCabund}. This has limited their usefulness in extragalactic science, since prohibitively large integration times are required at distances beyond 1Mpc.    

However, in \citet[][ hereafter Paper~I]{RSG_Jband}, we demonstrated that by isolating a spectral window in the $J$-band which is relatively free of molecular lines, abundances may be extracted at resolutions as low as $R\sim$3000. Not only does this dramatically reduce the required exposure times, but is also suitable for multi-object spectrographs. This then would allow entire galaxies at several Mpc to be abundance-mapped in only a few hours with an 8m-class telescope. In a further paper, we showed that this technique was ideally suited to extremely large telescopes (ELTs), with tremendous gains due to the large aperture and the adaptive optics system optimised for the NIR. We estimated that with the E-ELT we could expect to obtain abundances from individual RSGs out to distances of tens of Mpc in less that one night's integration \citep{Evans11}.

In \citet{PerOB1}, we performed a comprehensive test of this technique within a Solar metallicity environment, analysing a sample of stars in the Milky Way open cluster Per~OB1. We found an average metallicity $Z=-0.04 \pm 0.08$, consistent with the results from blue supergiants within the same cluster and in the Solar neighbourhoood \citep{Firnstein-Przybilla12,Nieva-Przybilla12}. We also demonstrated definitively that the technique works down to resolving powers less than $R = 3000$. In this paper, we undertake a similar analysis in two lower metallicity environments, the Small and Large Magellanic Clouds (SMC and LMC respectively), to test this technique for observations spanning a broad range of metallicities.

We begin in Sect.\ \ref{sec:obs} with a description of the data and its reduction. We describe the model grid we use, along with the analysis technique, in Sect.\ \ref{sec:anal}. The results are presented in Sect.\ \ref{sec:res}, and we compare our results to other studies of massive stars in Sect.\ \ref{sec:disc}. We conclude in Sect.\ \ref{sec:conc}.


\section{Observations and data reduction} \label{sec:obs}

\subsection{Observations}
We obtained observations of several stars in the LMC and SMC
using VLT+XSHOOTER \citep{XSHOOTER} under ESO programme number 088.B-0014(A) (PI
B.\ Davies). These observations provide continual spectral coverage from 0.3-2.4\um. The description of the observing strategy and the sample selection was described in \citet[][, hereafter D13]{rsg_teff}, which we briefly summarise here. The stars were observed in nodding ABBA mode with at least four exposures per star, and a randomized jitter at each position on the slit. For each of the instrument arms -- UVB, VIS, and NIR -- we used the 5.0\arcsec\ slits to minimize slit losses and obtain accurate spectrophotometry\footnote{The spectrophotometric aspect of these observations was not required in the present study, but was essential for the study presented in D13.}. The use of the broad slit meant that spectral resolution was determined by the seeing, which at 1\arcsec\ corresponds to a resolution $R \equiv \Delta \lambda / \lambda \simeq 5000$. The precise value of $R$ for each star was determined at the analysis stage (see Sect.\ \ref{sec:anal}), and was found to be within $R=4000-8000$. Integration times were chosen to achieve a signal-to-noise ratio (SNR) of at least 100 in the $J$-band (see Table 1 of D13). Flux standard stars were also observed each night and to correct for the atmospheric absorption in the NIR, telluric standard stars of spectral type late-B were observed within one hour of each science target. In general observing conditions were good on each night, and the seeing was around 1\arcsec\ or better. The standard suite of XSHOOTER calibration frames used in the data reduction process were taken at the beginning and end of the night \citep[for details see][]{Modigliani10}.

\subsection{Data reduction}
The initial steps of the reduction process were done using the XSHOOTER data reduction pipeline \citep[][]{Modigliani10}. These steps included subtraction of bias and dark frames, flat-fielding, order extraction and rectification, flux and wavelength calibration. The accuracy of the wavelength solution was determined by measuring the residuals of the
fitted arc lines with their vacuum wavelengths.  

The spectra of the science targets and the telluric standards were then extracted from final rectified two-dimensional orders. The strengths of the telluric lines were scaled in order to give the best cancellation across the diagnostic bandwidth (1.15-1.22\um). 


\section{Analysis}\label{sec:anal}

\subsection{Model grid}
For our project we have computed a new grid of model atmospheres. These atmospheres were generated using the MARCS code \citep{Gustafsson08}, which operates under the assumptions of LTE and spherical symmetry. The grid of models is four-dimensional, computed with a range of metallicities (\logz), gravities (\logg), effective temperatures (\teff) and microturbulent velocities ($\xi$). The chemical composition was scaled from Solar between [Fe/H]=-1.5 and +1.0 in steps of 0.25dex; \teff\ between 3400 and 4000K in steps of 100K, with further models at 4200K and 4400K; \logg\ between $-1.0$ and $+1.0$ in steps of 0.5 (in cgs units); and values of $\xi$ from 1 to 6\kms. All models were computed with an adopted stellar mass of $M_{\star}$=15\msun. Though RSGs may have masses between $\sim$8-25\msun, the only effect on the model of changing $M_{\star}$ is to alter the pressure scale height, which remains largely unchanged throughout the RSG mass range (see discussion
in Paper~I). The Solar-scaled abundance ratios were taken from \citet{GAS07}. The synthetic spectra were computed using the updated version of the SIU code, as described in \citet{Bergemann12}. SIU includes new opacities which are consistent with those in the DETAIL code \citep{DETAIL} which we use for statistical equilibrium calculations of Ti, Mg, Fe, and Si.

To increase the precision of our abundance measurements we have undertaken the task of computing corrections to our diagnostic lines which take into account departures from local thermodynamic equilibrium (LTE). Corrections to the \fei, \tii\ and \sii\ lines are presented in \citet{Bergemann12,Bergemann13}. Corrections to the \mgi\ lines are still in preparation, and are not considered in this work. The list of diagnostic lines used in this study are listed in Table \ref{tab:lines}. For this current study we have avoided using the \fei\ lines at 1.163826\um\ and 1.15936\um\ as these lines were subject to strong contamination from telluric absorption, making the line strengths uncertain.

\begin{table}[htdp]
\caption{Diagnostic spectral lines used in this analysis}
\begin{center}
\begin{tabular}{lcccr}
\hline \hline
Species & $\lambda$(\um) & $E_{\rm low} (eV)$ & $E_{\rm up} (eV) $ & $\log(gf)$\\
\hline
\fei & 1.188285 & 2.20 & 3.24 & $-$1.668 \\
\fei & 1.197305 & 2.18 & 3.22 & $-$1.483 \\
\sii & 1.198419 & 4.93 & 5.96 & 0.239 \\
\sii & 1.199157 & 4.92 & 5.95 & $-$0.109 \\
\sii & 1.203151 & 4.95 & 5.98 & 0.477 \\
\sii & 1.210353 & 4.93 & 5.95 & $-$0.351 \\
\tii & 1.189289 & 1.43 & 2.47 & $-$1.730 \\
\tii & 1.194954 & 1.44 & 2.47 & $-$1.570 \\
\hline
\end{tabular}
\end{center}
\label{tab:lines}
\end{table}%

\subsection{Continuum placement} \label{sec:continuum}
Our method is based on matching the observed line strengths relative to what we deem to be the continuum level. As described in Paper~I and in \citet{Evans11}, accurate continuum placement for RSGs is non-trivial. Their optical and near-IR spectra contain many thousands of blended molecular absorption lines, meaning that the true continuum may be well above the local maximum observed flux level. The spectral window that we adopt for this work was chosen specifically because of the low contamination by these lines, making continuum placement easier. Nevertheless there is still some weak molecular absorption present, and the `pseudo-continuum' we observe may be a few percent below than the true continuum. However, as long as the methodology we employ to determine the level of the pseudo-continuum is the same for both models and data, we will always find the same best-fitting model \citep[for further discussion, see][]{PerOB1}.

The first step of this process is to flatten out any large-scale slope that the data may have within the $J$-band window. To measure and remove this slope we first divide through by a template model spectrum which has had the strong absorption lines masked out. We use a low metallicity template, since these spectra have the cleanest continua free of molecular absorption\footnote{We investigated the effect on our results of using different templates, and found there was no effect provided the template had metallicity lower than [Z]=0.0, see Sect.\ \ref{sec:errors}.}. The resulting ratio spectrum was heavily smoothed with a low-order median filter, before fitting a 2nd-order polynomial. This polynomial is our measure of the low-order `tilt' of the observed spectrum, which is then removed by dividing through by the polynomial. 

To find the continuum level of this flattened spectrum, we ranked the pixels in order of flux values and determined the continuum to be the median of those with the top 20\%. This exact same methodology was applied to both data and models, to ensure that the placement of the pseudo-continuum was consistent. We investigated the robustness of our results to the exact method of continuum placement, experimenting with (for example) varying the cutoff value for selecting pixels to fit, varying the parameters of the template spectrum in the tilt-correction step (see above), as well as experimenting with alternative methods of continuum fitting such as those presented in \citet{Evans11}. We found that our results are not sensitive to this, provided the same method is applied to model and data, and the pixel cutoff is greater than $\sim$60\% (i.e.\ we use the top 40\% of ranked pixels) when determining the median continuum flux level. 

We also tested the sensitivity of our results to the signal-to-noise (S/N) of our data, by adding various levels of Gaussian noise. Since our method for continuum placement selects the highest value pixels, in the event of low S/N we would be selecting not the continuum but the high flux tail of the Gaussian noise. This would result in the continuum being placed too high, leading to the line strengths relative to the continuum being over-predicted, and ultimately an overestimate of the metallicity. We did indeed find that at low S/N the average abundance levels increased, but provided the S/N per pixel is greater than $\sim$60 then the measured abundances are stable at the level of $\pm0.02$dex. In practice, we have shown that it is possible to measure accurate abundances below this S/N limit by degrading the spectra to lower resolution at constant sampling of the resolution element, provided the degraded resolution is above $R=3000$ and the S/N per resolution element is $\ga 100$ \citep{PerOB1}.

\subsection{Spectral fitting methodology} \label{sec:fitting}
The basis for our analysis is that described in Paper~I. Briefly, we
compare the strengths of absorption lines in an input spectrum with those of a template spectrum  at each point in the model grid. The difference between the input and model spectra is used to compute a \chisq\ value, and the best fitting model is deemed to be that with the lowest \chisq. 

Before computing the \chisq\ value at each point in the grid, there are two subtle effects that must be taken into account which may produce spuriously high \chisq\ values. The first is that any relative velocity shift between model and data must be removed. This is done by an iterative cross-correlation procedure, with the model shifted until the measured offset is below 0.1 pixels. 

The second effect is that of a mismatch between the line-widths of the models and the data. Such a mismatch may be caused by variations in instrumental spectral resolution, due to e.g.\ instrument flexures or seeing variations, or by astrophysical effects such as macro-turbulence and stellar rotation. {Since the data presented here were taken with a wide 5\arcsec\ slit, seeing variations can cause the spectral resolution to vary from $4000 < R < 8000$}. In an equivalent width (EW) study we would not need to account for such factors as the flux within each line would be conserved. However, in the case of RSGs, broadening the spectral lines means that more of the unresolved molecular absorption that contributes to the pseudo-continuum is swallowed up by the diagnostic lines, meaning that the measured equivalent width actually increases as the resolution is degraded. 

To account for this, we again employ an iterative procedure. First, we take a grid of model spectra that have been degraded to a resolution of $R=10,000$, which we expect to be substantially higher than that of our data. We then find the model in this grid which has the best overall match to the diagnostic line {\it strengths}, i.e.\ the contrast between the the line-centre and the continuum. This quantity is obviously dependent on the spectral resolution, as for a given equivalent width this contrast will be greater for spectra of higher resolving power. We then measure the widths of the diagnostic lines in both the best-fit model and the observed data by fitting Gaussian profiles\footnote{The dominant source of line broadening in our data is instrumental, so Gaussian profiles are an appropriate choice for modelling the lines in our spectra.}. In this first iteration, we expect the line widths of the model to be narrower than the data, as we have deliberately overestimated the spectral resolution. We compare the difference in the fitted profile widths of the model and observed spectra, use these to estimate the true resolution of the data. This process is then repeated with a model grid that has been degraded to the updated estimate of the resolution, and we continue to iterate until the input and output resolution is stable. Applying this methodology we find that the estimated resolution converges to 5\% within four iterations. This corresponds to a systematic error in the inferred value of [Z] of $\pm0.04$dex, which we account for in our uncertaintiesÊ (see Sect.\ \ref{sec:errors}).

To check the sensitivity of our results to spectral resolution, we took the observed data and degraded it to lower resolutions, and re-performed the abundance analysis. We found that the best-fit parameters were stable down to measured resolutions of $R \ge 2000$, at which point the diagnostic lines begin to blend together. This is again consistent with our results for the RSGs in Per~OB1 \citep{PerOB1}. 

\begin{figure*}
\centering
\includegraphics[width=8.5cm]{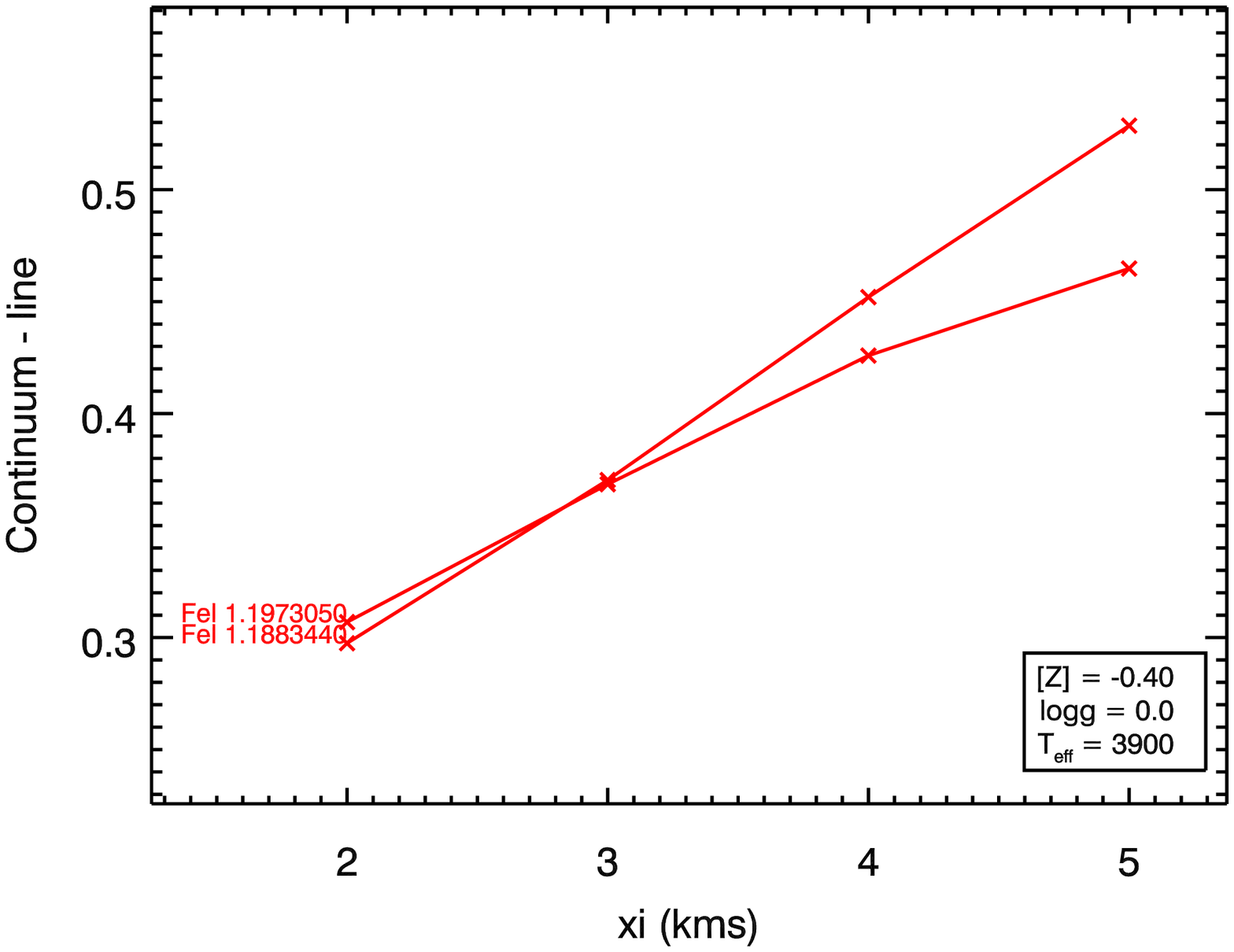}
\includegraphics[width=8.5cm]{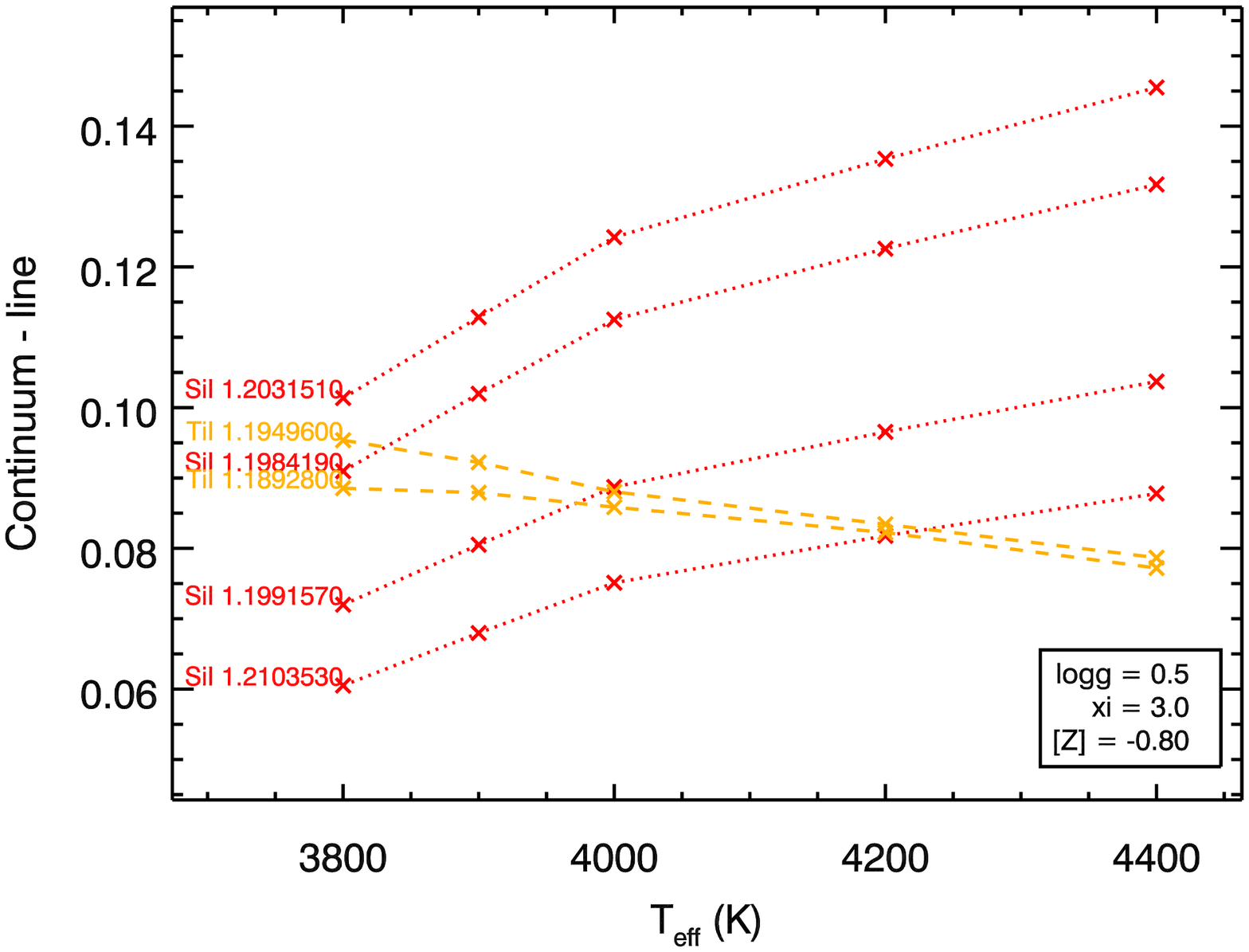}
\caption{An illustration of the effects of varying the model parameters on the line strengths. In the left panel, we plot the line strength ($\equiv$ continuum -- central line depth) for the strongest \fei\ lines as a function of $\xi$, demonstrating how the two lines behave differently. {\it Right panel:} the strengths of the \tii\ and \sii\ lines as a function of \teff. }
\label{fig:ew_behaviour}
\end{figure*}

\begin{figure*}
\centering
\includegraphics[width=17cm]{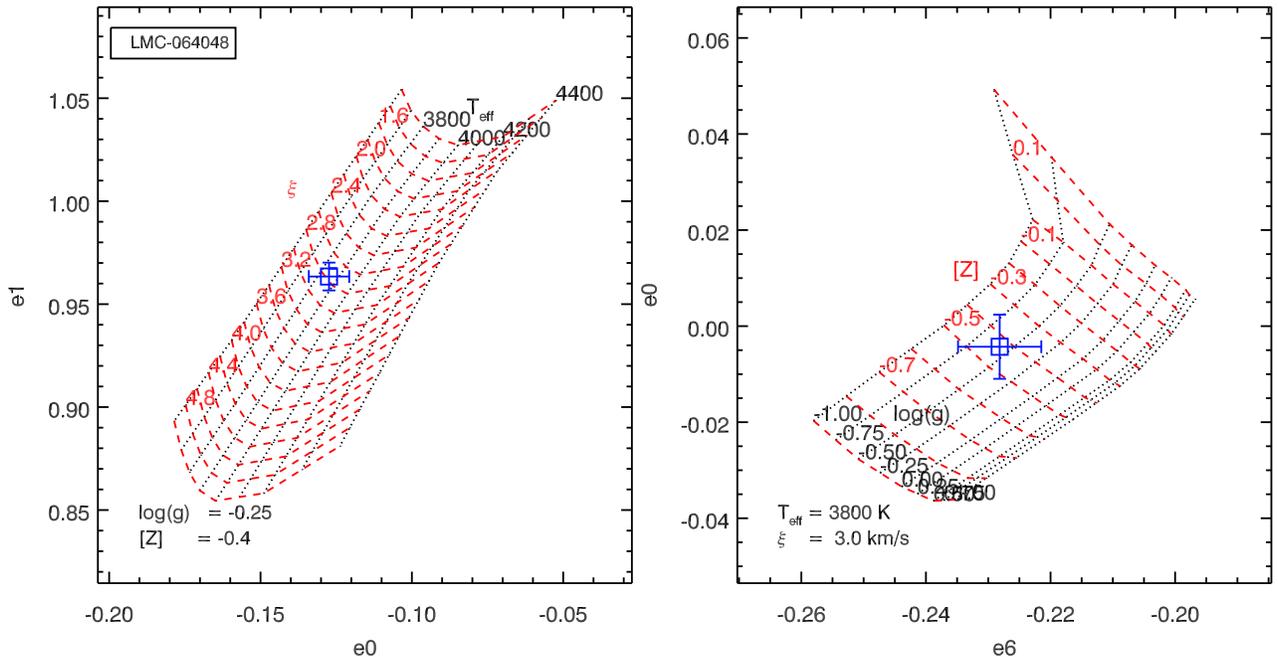}
\caption{Principle component analysis of the line strengths ($\equiv$ central line depth) in our model grid, illustrating the orthogonality between pairs of model parameters. In the left panel, we plot the first two principle components ($e0$ and $e1$) when \teff\ and $\xi$ are allowed to vary at fixed \logg\ and [Z]. For demonstration purposes we have taken the observed line-strengths of the first star in our sample, LMC~064048, and used the eigenvectors to calculate $e0$ and $e1$ (overplotted as a blue square). In the right panel we do the same, but for the other pair of model parameters. This time, the first pair of orthogonal components were $e0$ and $e6$. }
\label{fig:PCA}
\end{figure*}

\begin{figure}
\centering
\includegraphics[width=8.5cm]{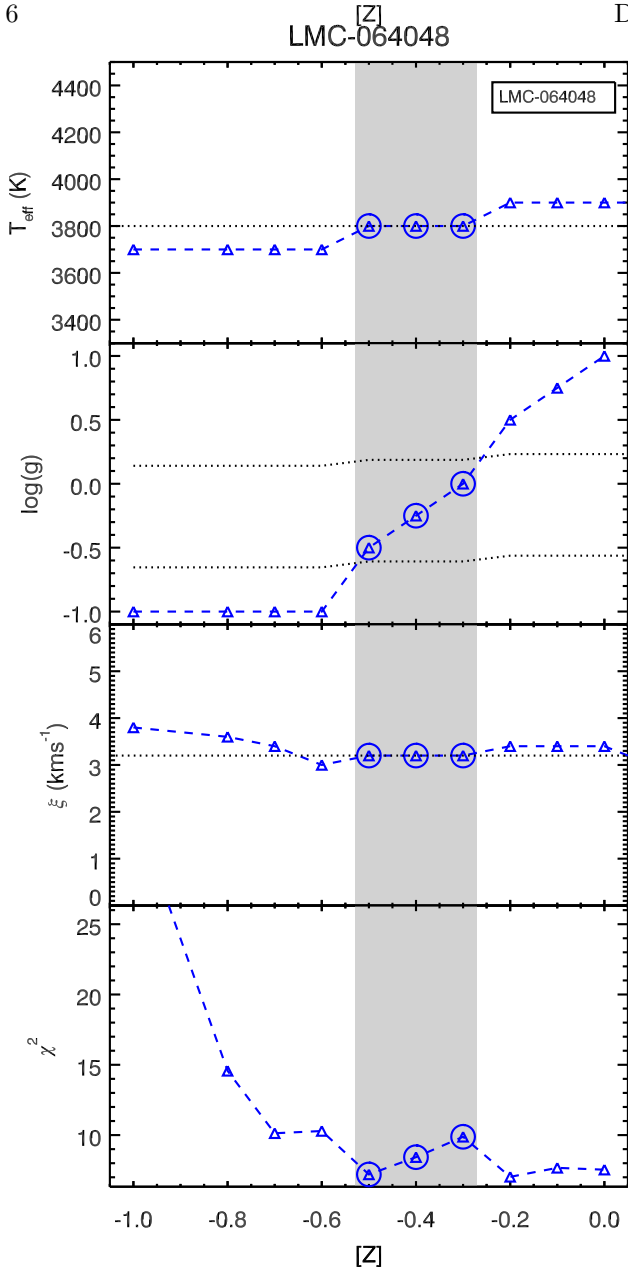}
\caption{The best-fitting values of \teff\ (top panel), \logg\ (second panel), and $\xi$ (third panel) for fixed values of [Z] for the star LMC~064048. The bottom panel shows the \chisq\ at each [Z]. The dotted lines in the second panel illustrate the maximum and minimum possible values of \logg, determined from the star's photometry and temperature. The circled points are those selected for determining the \chisq-weighted average, since it is these points which are within the possible gravity range (or closest to it). The dotted lines in the top and third panels denote the average \teff\ and $\xi$ respectively.  }
\label{fig:vary_z}
\end{figure}

\begin{figure*}
\centering
\includegraphics[width=17cm]{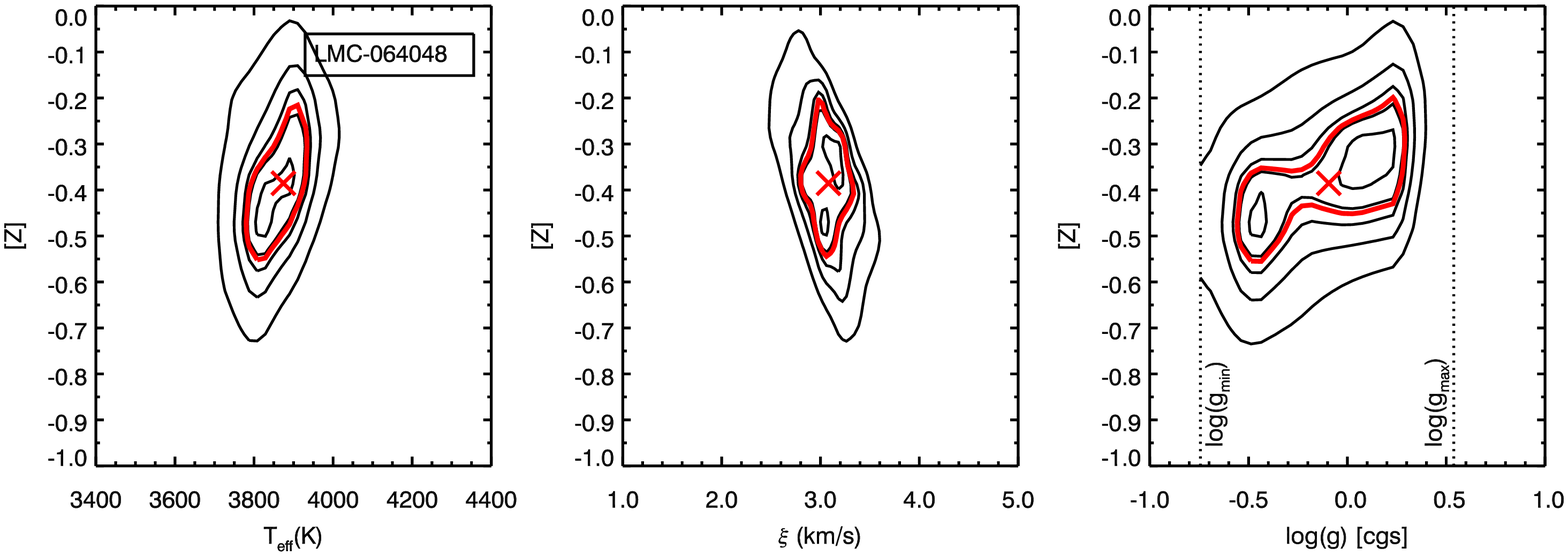}
\caption{Planes through the model grid showing the degeneracy between the parameters for the star LMC~064048, see text for how planes were created. The black contours indicate contours of equal \chisq\ values, and are drawn at levels of $\chi^2_{\rm min} + (1,2,3,5,10)$. The red contour is drawn at \chisq$= \chi^2_{\rm min} + 3$, which indicates our 1$\sigma$ uncertainty. The dotted lines on the right-hand plot indicate the minimum and maximum possible values of \logg\ given the star's temperature and luminosity. }
\label{fig:degen}
\end{figure*}

\subsection{Sensitivities of the diagnostic lines to the free parameters} \label{sec:sens}
Since there are four model parameters which can potentially influence the output spectrum, it is important to understand how each of these variables affects the relative strengths of the diagnostic lines we employ. Below we discuss the sensitivities of the diagnostic lines to each of the free parameters in our model grid. {We supply demonstrative figures (Figs.\ 1 - 3) to illustrate how certain lines respond to changes in one input parameter while the others are fixed, however we note that our fitting methodology is to consider all lines and all parameters simultaneously (see Sect. \ref{sec:errors}).}

\subsubsection{Microturbulence, $\xi$} \label{sec:xi}
The effect of this parameter is to increase the EW of stronger saturated lines, whilst the EW of weaker lines will be less affected. We therefore expect that the relative strengths of lines of the same element and ionization stage will be sensitive to this parameter. This can be simply demonstrated by studying the variation in line strengths as $\xi$ is increased while keeping the other parameters fixed, where we see the two \fei\ lines responding differently (\fig{fig:ew_behaviour}, left panel). {We note that the crossing of the two curves in the left part of Fig. 1 is a subtle non-LTE effect. Both lines belong to the same multiplet and, thus, in LTE the curves displaying their strengths would not cross. However, in our non-LTE calculations the multiplet transitions are treated as individual lines and have distinct departure coefficients which respond differently when the atmospheric parameters are varied.}

\subsubsection{Effective temperatures \teff} \label{sec:teff}
Spectra of RSGs in the $J$-band window contain lines that have different excitation potentials \citep[Table \ref{tab:lines}, see][]{Bergemann12,Bergemann13}. Therefore, we may expect that the ratio of (for example) the \sii\ to \tii\ line strengths would be sensitive to \teff. This can be demonstrated by looking at how the strengths of these lines change as a function of \teff\ with the other parameters fixed (\fig{fig:ew_behaviour}, right panel).

\subsubsection{Gravity and metallicity} \label{sec:z-grav}
Separating the effects of these two parameters is less straight-forward than for \teff\ and $\xi$. Overall, increasing (decreasing) metallicity (gravity) at fixed \teff\ and $\xi$ has the effect of increasing the strengths of all the lines, though some lines are more sensitive than others. This means that there is a degree of degeneracy between these two parameters. 

To investigate this further, we performed principle component analysis (PCA) on the matrix of the strengths of the eight diagnostic lines. We did this in two stages -- first, we extracted two dimensional grids of line-strengths for each line as a function of \teff\ and $\xi$, at fixed \logg\ and [Z]. We computed the eigenvectors, and searched for pairs of eigenvectors which were orthogonal, preferring the earliest possible components since these account for the greatest variance. We then repeated this process allowing \logg\ and [Z] to vary at fixed \teff\ and $\xi$. 

We show an example of this in \fig{fig:PCA}. The left panel shows the first pair of orthogonal eigenvectors when allowing \teff\ and $\xi$ to vary. Here, the first two eigenvectors $e0$ and $e1$ display a large degree of orthogonality, and which account for 99.8\% of the variance. The interpretation of this is that these two parameters are readily constrained from the very different effect that each has on the strengths of the diagnostic lines. Overplotted on this panel is the datum for the first star in our sample, LMC~064048, for which we have computed the projection onto the $e0-e1$ plane using the star's line strengths. The very small error bars illustrate that \teff\ and $\xi$ can be constrained to high precision. 

In the right panel of \fig{fig:PCA} we show a similar analysis but for varying \logg\ and [Z]. Here it can be seen that to find orthogonal eigenvectors we had to go to seventh eigenvector $e6$, which accounts for only a very small fraction of the variance ($\ll$1\%). This is illustrated by the error bars on the observation of LMC~064048, which are much larger relative to the parameter space covered by the models. The interpretation of this is that the degeneracy between \logg\ and [Z] is harder to break due to the subtle differences that exist between their effects on the strengths of the spectral lines, and detecting these differences requires high signal-to-noise data. 

Since obtaining such high S/N data may be challenging, especially when considering that there may be systematic errors in both the data reduction and the spectral synthesis that we have not accounted for, we explored an alternative method to constrain \logg\ and hence better understand the degeneracy between this parameter and [Z]. First, the model grid was subsampled in flux onto a finer grid, with 0.2\kms\ steps in $\xi$, 0.1dex steps in [Z], and 0.25dex steps in \logg. The strengths of the diagnostic lines were measured for each model in the grid. For each subsampled value of [Z], we extracted a three-dimensional sub-grid (\teff, $\xi$, and \logg), and performed a \chisq-minimization search to find the best fitting values of the three free parameters at that value of [Z]. Since the best-fitting \teff\ and $\xi$ are well constrained for a given set of line-strengths (see above), this result then tells us the best-fitting value of \logg\ for that particular value of [Z]. We then repeat this for all values of [Z].

An example of the results of this analysis can be found in \fig{fig:vary_z}, which shows the best-fit values of \teff, \logg\ and $\xi$ for a fixed value of [Z], again for the first star in our sample. The plot further helps to illustrate the degeneracies in our fitting method. The best-fitting values of the parameters \teff\ and $\xi$ are stable regardless of what value [Z] is fixed at, and hence that there is very little degeneracy between them. 

The second-top panel demonstrates the degeneracy between [Z] and \logg. We see that for low values of [Z], lower gravities are preferred. This is because lowering [Z] reduces the line strengths, and so to match the line strengths of the data we must select low values of \logg\ to compensate. Similarly, at high metallicities, we must select the highest gravity models to best match the line strengths. At intermediate metallicities we see a transition from low to high gravities. 

However, we can put prior constraints on \logg. For a given \teff\ and luminosity (the latter obtained from D13) we can estimate the radius $R$ of the star. Then, by selecting an appropriate mass range $M$ for that value of $L$ using evolutionary models \citep[8-30\msun, e.g.][]{Mey-Mae00}, we can estimate the surface gravity from $g = GM/R^2$. Allowing for a contribution from convective pressure which would reduce the {\it effective} gravity by up to 0.3dex \citep{Chiavassa11,PerOB1}, we can identify the extreme possible values of \logg. When taking the \chisq-weighted mean of each of the parameters within this range, or simply just their values at the \chisq\ minimum, we find very similar results to those in the by the PCA analysis (see above).


\begin{figure*}
\centering
\includegraphics[width=15cm]{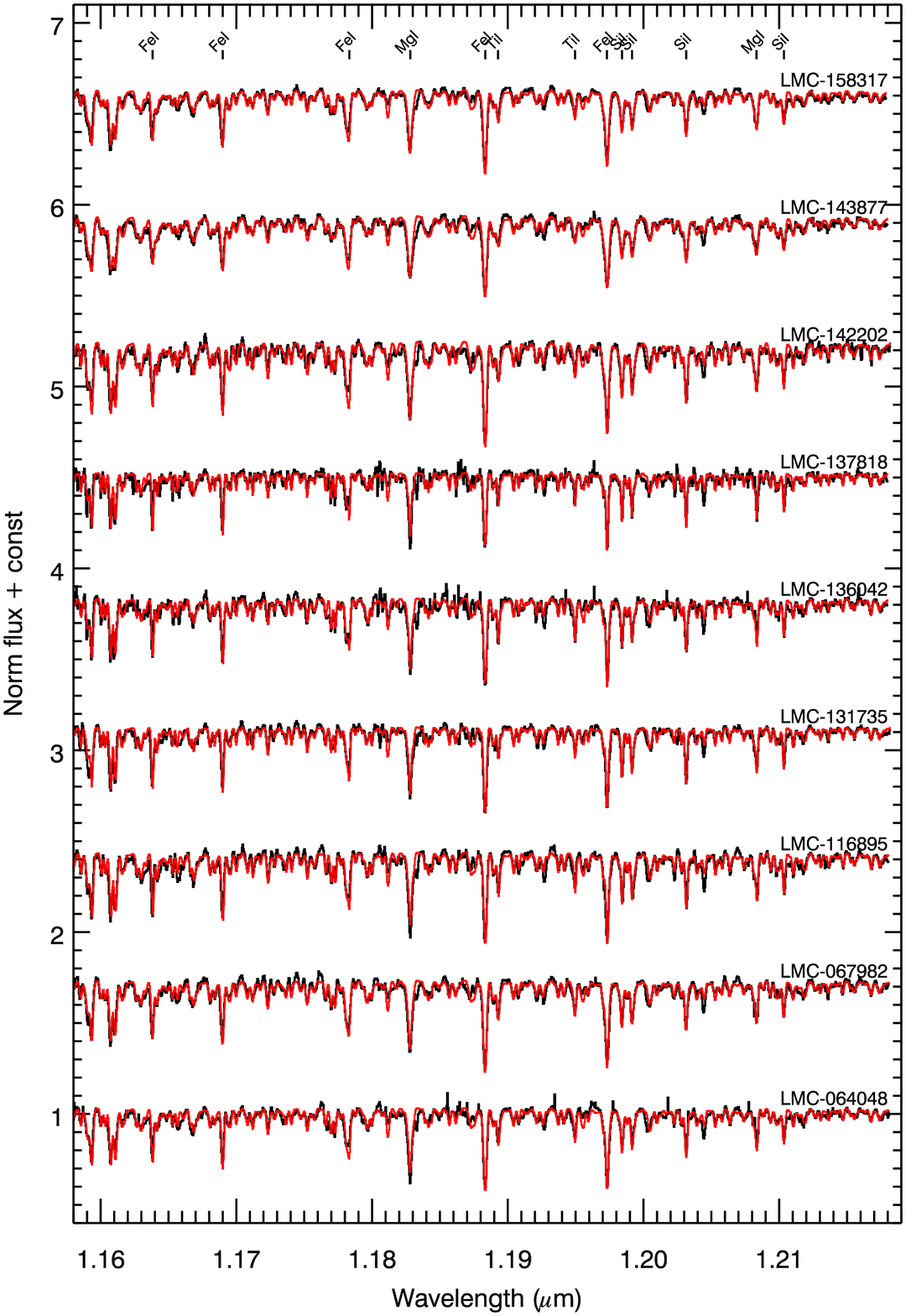}
\caption{The spectra of all LMC stars observed in the $J$-band window, along with best-fit spectra. The two \mgi\ lines, for which we do not yet have non-LTE corrections, have been excluded when fitting the spectra}
\label{fig:fits}
\end{figure*}
\begin{figure*}
\centering
\includegraphics[width=15cm]{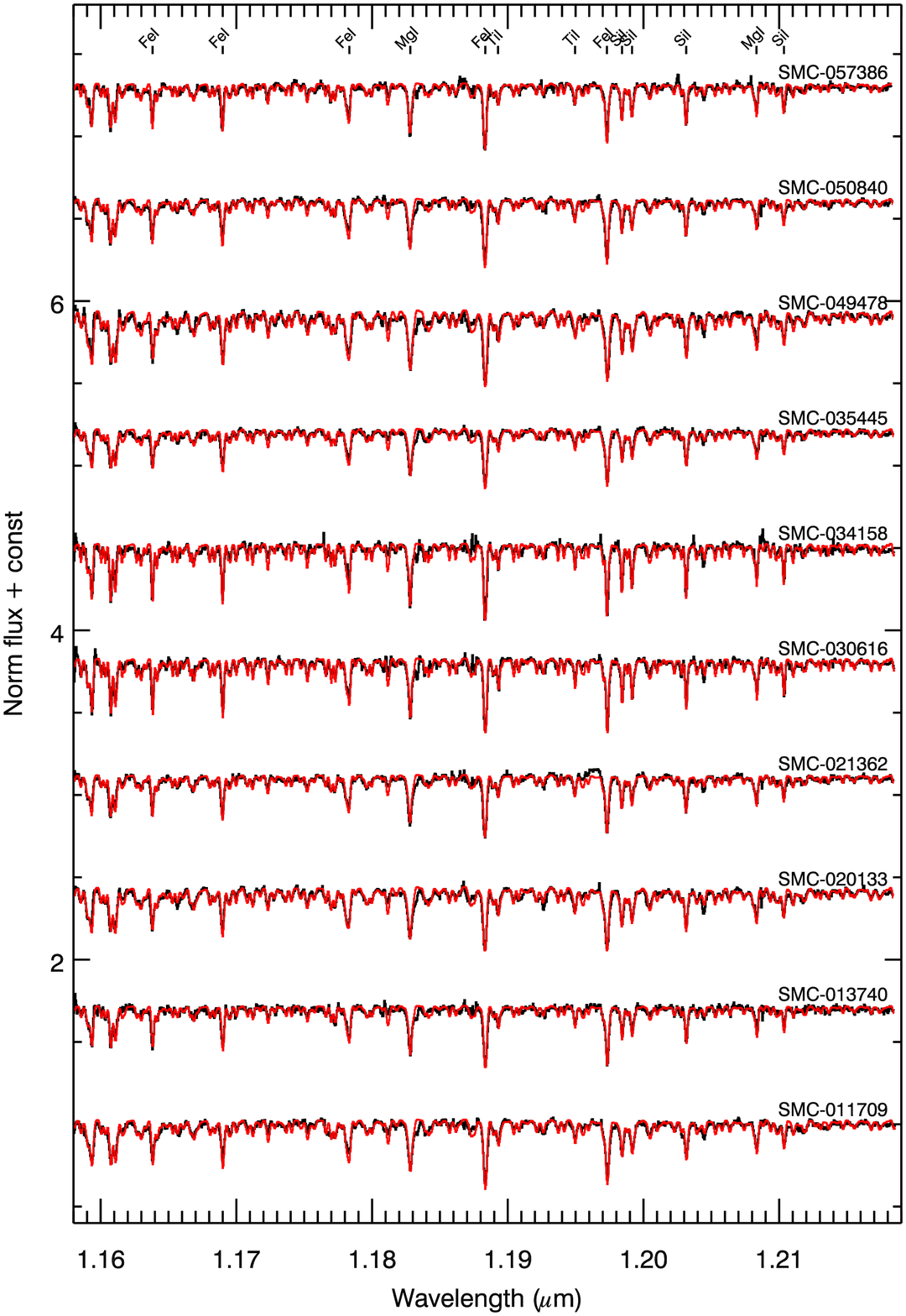}
\caption{Same as \fig{fig:fits}, but for the SMC stars.}
\label{fig:smcfits}
\end{figure*}

\subsubsection{Best-fit parameters and their uncertainties} \label{sec:errors}
Our methodology for finding the best-fitting model is as follows. Using the subsampled grid (see above), we compare the line-strengths of the observed spectrum $\mathcal{O}$ with those in the model grid $\mathcal{M}$ at the best-fitting resolution. At each point in the grid we compute the un-reduced \chisq\ value from the eight diagnostic lines, 

\begin{equation}
\chi^2 = \sum\limits_{i}^{n} (\mathcal{O}_i - \mathcal{M}_i)^2 / \sigma_i^2  
\end{equation}

\noindent where $\sigma$ is the error on the line-strength, determined by the S/N, and which does not account for the error in the spectral resolution which is treated separately (see below). We ignore all points in the grid where the value of \logg\ is unphysical. Since we know the luminosities of the stars in our sample, this also allows us to rule out grid points with low-\teff\ / high-\logg\ (and vice versa), since stars at these grid points would have anomalously high (or low) masses. The best-fitting values of each parameter are determined from the weighted mean of all points in the grid, where the weight $w$ is given by $w_i = \exp(-\chi_i^2)$. Eliminating unphysical regions of parameter space before computing the averages is useful for eliminating localised \chisq-minima, which may skew the best-fits for the parameters. 

When determining the uncertainties on each of the four free parameters, we consider three sources of random errors. We discuss each of these below individually. The total errors are taken to be the quadrature sum of these three sources of uncertainty.

\paragraph{`Degeneracy' errors:}
This is the term we use to describe the uncertainty caused by several points within the model grid having \chisq\ values which are close to that of the best-fitting model. To determine this error, we define a \chisq\ tolerance, $\Delta \chi^2 \equiv \chi^2_{\rm min} + 3$, where \chisq$_{\rm min}$ is the minimum {\it unreduced} \chisq. The tolerance value of 3 is chosen for the following reasons. If all line strengths in the observed spectrum are on average matched to 1$\sigma$ by those in the model spectrum, the unreduced \chisq\ value should be equal to the number of diagnostic lines $n$, which in this case $n=8$. Should one of the lines be fitted to only 2$\sigma$, the rest being fitted to 1$\sigma$, the unreduced \chisq\ becomes $(n-1)\times1^2 + 1\times4^2=n+3$ \footnote{This tolerance threshold for \chisq\ compares well to the `classical' value of $\Delta$\chisq=2.3 for a 1$\sigma$ deviation from the peak of a purely Gaussian probability distribution.}. We consider all models which are able to fit the data to better than $\Delta \chi^2$ to define the limits of the degeneracy errors.

To illustrate these errors, we calculate the \chisq\ values at every point in the grid and take two-dimensional projections. For one pair of parameters, say \teff\ and [Z], we determine the minimum \chisq\ for every value of \teff\ and [Z] whilst allowing the other two parameters (in this case \logg\ and $\chi$) to be free, other than the limits we have already placed on \logg\ (see Sect.\ \ref{sec:z-grav}). We thus construct a 2-D plane of minimum \chisq\ values as a function of \teff\ and [Z]. On each of these planes, we then draw contours of equal \chisq. An example of this is shown in \fig{fig:degen}. This figure shows again how tightly constrained \teff\ and $\xi$ are, whilst also demonstrating again the degeneracy between [Z] and \logg. Typical errors on [Z] from this source of uncertainty are between $0.1-0.2$dex.

\paragraph{`Resolution' errors:} 
As discussed in Sect.\ \ref{sec:fitting}, part of the fitting involves determining the spectral resolution, which we do by matching the line widths at the observed line depths. We do this by applying an iterative procedure until convergence at the 5\% level is achieved. Were we basing our fitting methodology on line equivalent widths then this uncertainty would be combined with the degeneracy errors. However, since we are fitting both the line depths and the line widths, we consider these sources of error separately.

To determine the magnitude of this uncertainty on our results, we re-ran the analysis above but compared the observed line strengths to those of models which had been degraded to the observed resolution $\pm$5\%. The effect on \teff\ and $\xi$ was minimal, since these parameters are sensitive to line depth {\it ratios}, whereas altering the resolution uniformly alters only the line depths. The effect on \logg\ was also small compared to the degeneracy errors, typically $\la$0.1dex. The effect on [Z] was again minor, $\pm$0.04 on average, but large enough that they cannot be neglected.

\paragraph{Continuum placement errors:}
These errors refer to the uncertainties described in Sect.\ \ref{sec:continuum}, which concern the placement of the pseudo-continuum. As has already been discussed, this is not the true continuum since unresolved molecular absorption reduces the overall maximum flux level. However, as long as the same principles and methodology are applied to both models and observations then the line to psuedo-continuum ratio is still an accurate measure of metallicity \citep[see also discussion in][]{PerOB1}. 

To investigate the effect of this uncertainty we re-ran the analysis several times, each time fitting the continuum in a slightly different way. We varied aspects such as the metallicity of the template spectrum (used to correct for any spectral tilt), the fraction of pixels used to determine the highest flux level, as well as alternative continuum placement methods we have experimented with previously, such as those described in \citet{Evans11}. Each time the difference in the fitted parameters was minimal, with the abundances stable to within $\pm$0.02dex. We conclude that this source of error is small compared to the other two described above, but is included nonetheless.

\begin{table*}
\caption{Best fit model parameters for each star in this study.}
\begin{center}
\begin{tabular}{lccccc}
\hline \hline
Star & \teff\ (K) & log(g) & $\xi$ (km/s) & [Z] & $R (\pm 5\%)$ \\
\hline
               LMC-064048 & 3860 $\pm$  70 & -0.2 $\pm$ 0.4 & 3.1 $\pm$ 0.2 & -0.42 $\pm$ 0.17 & 5600 \\
               LMC-067982 & 3910 $\pm$  60 & -0.3 $\pm$ 0.3 & 3.7 $\pm$ 0.3 & -0.43 $\pm$ 0.14 & 5400 \\
               LMC-116895 & 3950 $\pm$  60 & -0.3 $\pm$ 0.5 & 3.2 $\pm$ 0.2 & -0.30 $\pm$ 0.18 & 6400 \\
               LMC-131735 & 4110 $\pm$  50 & -0.0 $\pm$ 0.2 & 3.3 $\pm$ 0.2 & -0.50 $\pm$ 0.09 & 6700 \\
               LMC-136042 & 3850 $\pm$  50 &  0.1 $\pm$ 0.1 & 2.7 $\pm$ 0.2 & -0.07 $\pm$ 0.07 & 6900 \\
               LMC-137818 & 3990 $\pm$  50 &  0.0 $\pm$ 0.2 & 2.3 $\pm$ 0.2 & -0.50 $\pm$ 0.13 & 8200 \\
               LMC-142202 & 4100 $\pm$  50 & -0.5 $\pm$ 0.3 & 4.2 $\pm$ 0.2 & -0.28 $\pm$ 0.08 & 5100 \\
               LMC-143877 & 4060 $\pm$  80 & -0.3 $\pm$ 0.5 & 3.7 $\pm$ 0.3 & -0.29 $\pm$ 0.17 & 4000 \\
               LMC-158317 & 4040 $\pm$  80 & -0.0 $\pm$ 0.5 & 3.6 $\pm$ 0.4 & -0.37 $\pm$ 0.19 & 4600 \\
	{\it Average, LMC} & & & & {\it -0.37 $\pm$ 0.14} \smallskip \\
               SMC-011709 & 4020 $\pm$  50 & -0.3 $\pm$ 0.1 & 3.0 $\pm$ 0.2 & -0.53 $\pm$ 0.10 & 4900 \\
               SMC-013740 & 3970 $\pm$  70 &  0.3 $\pm$ 0.5 & 2.6 $\pm$ 0.2 & -0.59 $\pm$ 0.22 & 5800 \\
               SMC-020133 & 3970 $\pm$  80 & -0.2 $\pm$ 0.4 & 2.3 $\pm$ 0.3 & -0.21 $\pm$ 0.20 & 5000 \\
               SMC-021362 & 3970 $\pm$  70 &  0.0 $\pm$ 0.5 & 2.7 $\pm$ 0.2 & -0.57 $\pm$ 0.20 & 5400 \\
               SMC-030616 & 4040 $\pm$  60 & -0.2 $\pm$ 0.5 & 2.8 $\pm$ 0.2 & -0.47 $\pm$ 0.17 & 7000 \\
               SMC-034158 & 4180 $\pm$  70 &  0.1 $\pm$ 0.5 & 3.2 $\pm$ 0.2 & -0.48 $\pm$ 0.15 & 7300 \\
               SMC-035445 & 4040 $\pm$  80 & -0.0 $\pm$ 0.5 & 2.4 $\pm$ 0.2 & -0.53 $\pm$ 0.21 & 5200 \\
               SMC-049478 & 4110 $\pm$  60 & -0.3 $\pm$ 0.4 & 3.2 $\pm$ 0.2 & -0.24 $\pm$ 0.12 & 4900 \\
               SMC-050840 & 3920 $\pm$  50 & -0.3 $\pm$ 0.3 & 3.0 $\pm$ 0.2 & -0.72 $\pm$ 0.12 & 5400 \\
               SMC-057386 & 4120 $\pm$  50 & -0.0 $\pm$ 0.5 & 2.9 $\pm$ 0.2 & -0.57 $\pm$ 0.15 & 5900 \\
	{\it Average, SMC} & & & & {\it -0.53 $\pm$ 0.16} \smallskip \\

\hline
\hline
\end{tabular}
\end{center}
\label{tab:results}
\end{table*}%

\begin{figure}
\centering
\includegraphics[width=8.5cm]{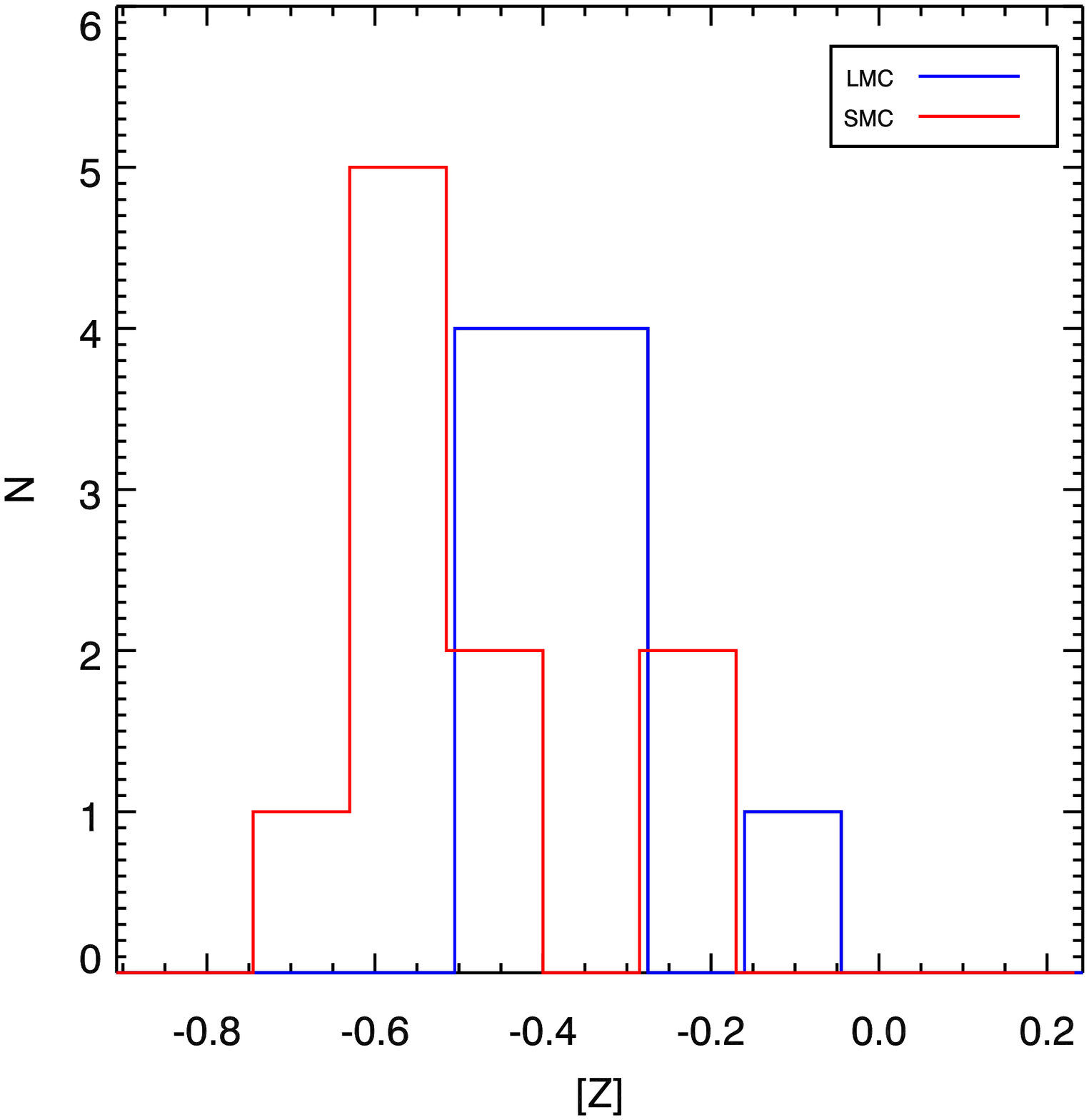}
\caption{Metallicity distributions of the RSGs in the two Magellanic Clouds. }
\label{fig:hist}
\end{figure}

\section{Results of the $J$-band analysis} \label{sec:res}

The data, along with the best fitting model spectra, are shown in Figs.\ \ref{fig:fits} and \ref{fig:smcfits}. While the match to the diagnostic lines is excellent, the fits to the unresolved features of the pseudo-continuum is also very good. Since the latter were not used to diagnose the best fitting model parameters, this gives us further validation that our models are performing well. The only features not well matched are the two \mgi\ lines, and an unknown feature at 1.2045\um. The explanation for the former is likely non-LTE effects, the corrections for which will be presented in a forthcoming paper (Bergemann et al., in prep), while we currently do not have an explanation for the latter. 

The best-fitting model parameters are listed in Table \ref{tab:results}. As described in Sect.\ \ref{sec:sens}, the parameters \teff\ and $\xi$ are sensitive to line {\it ratios}, and with high S/N data in which the lines are well resolved, these parameters are strongly constrained. As discussed in the previous sections and illustrated in Figs.\ \ref{fig:vary_z} and \ref{fig:degen}, \teff\ and $\xi$ are stable to well within $\sim$10\% regardless of what value we fix the metallicity to. The major source of error in the metallicity $[Z]$ is the degeneracy with \logg, which accounts for an uncertainty of approximately $\pm$0.1dex. 

\begin{figure*}
\centering
\includegraphics[width=17cm,bb=0 0 880 396,clip]{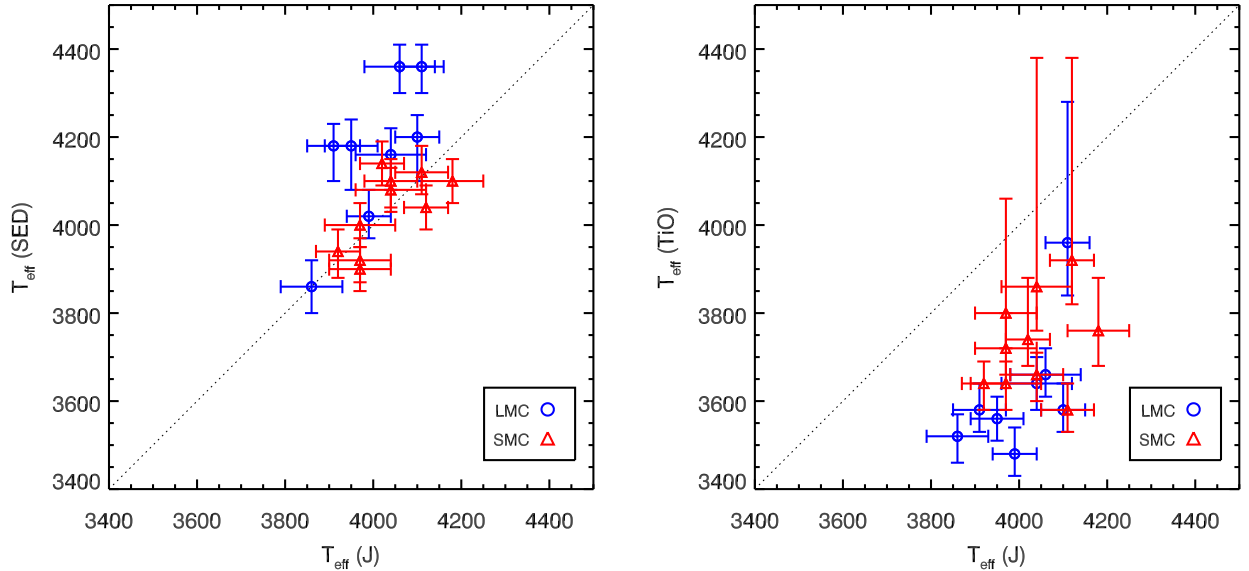}
\caption{Comparison of the temperatures derived in this study, \teff($J$), with those from fits to the stellar SEDs (left panel) and from fits to the TiO bands in the optical (right panel).  }
\label{fig:teff-sed}
\end{figure*}

For the LMC stars, we find an average metallicity of $[Z]=-0.37\pm0.14$, consistent with other measurements of young stars in this galaxy (see following section). The one object which is slightly discrepant with the others is LMC~136042 which has a metallicity 0.3 dex higher than the average. As discussed in D13, this star's spectrum is blended with a nearby hot star, evident from the blue region of the spectrum. This companion should not contribute much to the $J$-band flux, indeed if it did then we would expect the diagnostic lines to be diluted and for the fitted metallicity to be abnormally {\it low}. Therefore, we do not see a good reason to discard this object from our sample. 

\begin{figure}
\centering
\includegraphics[width=8.5cm]{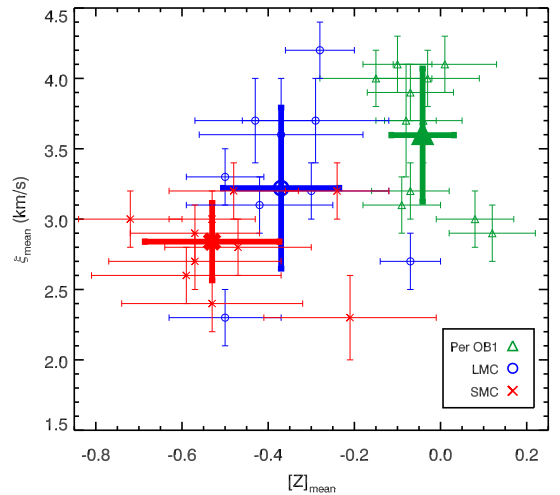}
\caption{The best-fit microturbulent velocities as a function of metallicity. The average values and their standard deviations for the three environments -- Per OB1 \citep{PerOB1}, and the LMC and SMC (this work) --  are indicated by the bold symbols and thicker error bars. }
\label{fig:abun_turb}
\end{figure}

In terms of the SMC stars, we find an average metallicity of $[Z]=-0.53\pm0.16$, slightly on the higher side of other measurements of young stars, but consistent to within the errors (see following section). The distribution of metallicities is peaked at $\sim$-0.6dex but with a high metallicity tail (see \fig{fig:hist}). We also find that the average value of $\xi$ is slightly lower in the SMC than for the LMC and for the Per~OB1 stars studied in \citet{PerOB1} ($\xi_{\rm SMC} = 2.8\pm0.3$\kms, as opposed to $3.3\pm0.5$\kms\ and $3.6\pm0.5$\kms\ for the LMC and Per~OB1 respectively, see \fig{fig:abun_turb}). {Though there does seem to be a weak trend of decreasing $\xi$ at lower metallicities, this is not borne out by a study of RSGs in NGC~6822, a galaxy found to have similar metallicity to the SMC \citep{NGC6822}.}


Finally, we note that we do not see any systematic trends with the fitted parameters, for example $R$ and [Z], which would indicate obvious systematic errors.

\subsection{Comparison of temperatures with SED fits}
In D13 we showed that the measurements of RSG effective temperatures using the TiO bands were unreliable, and presented a more robust way of measuring RSG temperatures using their optical-NIR spectral energy distributions (SEDs). Since the sample of the D13 is identical to that used here, it is natural to investigate how the effective temperatures measured here from the diagnostic $J$-band lines compare to those measured from SED fits.

In \fig{fig:teff-sed} we plot the \teff\ measurements of the current study against those from the SED fits of D13. The means and standard deviations of the quantity $\Delta T (\equiv T_{\rm J} - T_{\rm SED})$ are $160\pm110$K and $48\pm68$K for the LMC and SMC respectively.  This tells us that overall our MARCS models with non-LTE corrections are able to simultaneously match the optical-NIR continuum as well as the strengths of the atomic lines in the $J$-band. {Though the agreement between the two methods is excellent for the SMC stars, there is a systematic offset for the LMC stars, which for the two stars with the highest SED temperatures is 3-4$\sigma$}. At the present time we have no explanation for this, or whether indeed this offset is significant. A planned study of stars at higher and lower metallicities will address this. 

For completeness, in \fig{fig:teff-sed} we also plot the temperatures from the $J$-band fits against those derived from the TiO band strengths, also from D13. We find little correlation between these two, with the TiO temperatures being on average 300\,K cooler, mirroring the findings in D13.

Since there is a small discrepancy between the $J$-band effective temperatures and those determined from SED fits, it is natural to ask what systematic effect this may have on the derived metallicities. To answer this question we reanalysed the spectra in the same way as in Sect.\ \ref{sec:fitting}, but instead of allowing temperature to be a free parameter we fixed it to be the same as the SED temperatures from D13. The fitted metallicities of each star was typically within 1$\sigma$ of those quoted in Table \ref{tab:results}, while the mean metallicities of the LMC and SMC, which now were found to be -0.36 and -0.50 respectively, were stable to within $\pm$0.03dex.

\section{Discussion: a comparison with other metallicity studies of the Magellanic Clouds} \label{sec:disc}

The average metallicities of the two galaxies are listed in Table \ref{tab:results}. For the LMC we find $[Z]=-0.37 \pm 0.14$, and for the SMC we find $[Z]=-0.53 \pm 0.16$. These are relative to an adopted Solar value of \zsun=0.012 \citep{GAS07}. 


Below we discuss these results in the context of other metallicity studies of the Magellanic Clouds. In this discussion we restrict ourselves to studies of young ($\la$100Myr) stars, since these should be the most comparable to the results presented here. Compilations of these studies are presented in Tables \ref{tab:lmc} and \ref{tab:smc}. We also note that in our abundance analysis we have kept all metals scaled to their Solar values relative to each other. We will explore the possibilities of non-Solar $\alpha$-to-iron ratios at the end of this Section.

\subsection{Abundance studies of the LMC}
The most recent comparable studies to that presented here are those of \citet{Dufton06} and  \citet{Trundle07}. These authors studied samples of B stars in clusters and associations in both Magellanic Clouds, using non-LTE spectral analysis\footnote{The Fe lines were treated in LTE in \citet{Trundle07}, though non-LTE effects are thought to be small in the parameter range occupied by the stars in their sample \citep{Thompson08}. }. For the LMC, they studied stars in two clusters, N11 and NGC~2004. In general they found abundances of Fe and O which were {-0.3dex} with respect to Solar, and abundances of Si and Mg which were lower by $\sim$0.1dex relative to H. A sample of F~supergiants were the subject of an LTE study by \citet{Hill95}, finding [O,Mg,Fe] abundances $\sim$-0.25dex relative to H, while  Si was super-Solar. A re-analysis of the same stars by \citet{Andrievsky01} found an non-LTE-corrected [O] abundance of -0.16dex, and \hbox{[Fe]=-0.34} (LTE). These results are consistent with \citet{Hill95}, but serve as an illustration of the uncertainties. Two separate studies of cepheids found abundances in the range -0.1 to -0.2dex \citep{Luck98,Romaniello08}. 

In general, these previous studies converge at an average metal abundance [Z] of -0.3$\pm$0.1. Our average value from our sample of RSGs is therefore perfectly consistent with these earlier results.

\subsection{Abundance studies of the SMC}
For this galaxy there is a larger degree of disparity between estimates of the metallicity from a variety of abundance probes. In Table \ref{tab:smc} we summarise selected abundance determinations from young stars in the SMC. These studies have looked at abundances of individual elements rather than the global metallicity $Z$. We have concentrated on the elements O, Si, Mg and Fe (where available) since these elements should be relatively unaffected by stellar evolution for all but the most evolved stars \citep{Brott11}. 

For studies of hot stars, we consider only those which include the effects of non-LTE, which are known to be non-negligible in hot star atmospheres. Studies of main-sequence B stars in the young cluster NGC~330 \citep{Korn00,Lennon03,Trundle07} have found O abundances in the range -0.4dex to -0.8dex relative to the Solar value of $\rm \log(O/H) + 12 = 8.66$ \citep{GAS07}. Since one would expect the abundances in a young cluster to be uniform this serves to illustrate the uncertainties in this type of work. Abundances of Mg and Si range from -0.7 to -0.9dex compared to Solar. A study of another young cluster, NGC~346, and the isolated B star AV~304, found abundances for these three elements that were consistent with the NGC~330 studies (-0.5 to -0.8dex for O, Si and Mg). Analysis of {\it evolved} hot stars, B supergiants \citep{Trundle05} and A supergiants \citep{Venn95,Venn99}, again revealed similar values, though the latter is an LTE study. 

Previous analysis of {\it cool} evolved massive stars have not included non-LTE effects, but are included here for completeness, and are discussed within the context of our recent investigations into non-LTE corrections for cool supergiants \citep{Bergemann12,Bergemann13}. Another study of NGC~330, this time of the K supergiants, revealed abundance levels which were slightly lower than those obtained from analysis of hot stars, with abundances of O, Si, Mg and Fe all between 0.8-0.9dex below Solar \citep{Hill99}. This could in part be explained by non-LTE effects: \citet{Bergemann12} showed that for certain Fe lines the assumption of LTE would result in abundance underestimates $>0.1$dex at temperatures of $>$4400K, though the effect seemed to be the opposite for Si lines \citep{Bergemann13}. The analysis of K supergiants in the SMC field  were higher, between 0.5-0.8dex below Solar \citep{Hill97a,Hill97b}. 

Finally, from analysis of Cepheids, the abundances of O, Si, Mg and Fe were found to be 0.5-0.6dex sub-Solar \citep{Luck98}, though a more recent study by \citet{Romaniello08} found an iron abundance of -0.7dex relative to the Solar value of $\rm \log(Fe/H)+12 = 7.45$. Both studies assumed LTE. 

To summarise, young star abundance determinations seem to have a somewhat large spread, typically finding between 0.5-0.8dex below Solar. In this context, our result of $[Z] = -0.53 \pm 0.16$ is at the upper boundary of these other results, but consistent within the 1$\sigma$ errors. 

\begin{figure}
\centering
\includegraphics[width=8.5cm]{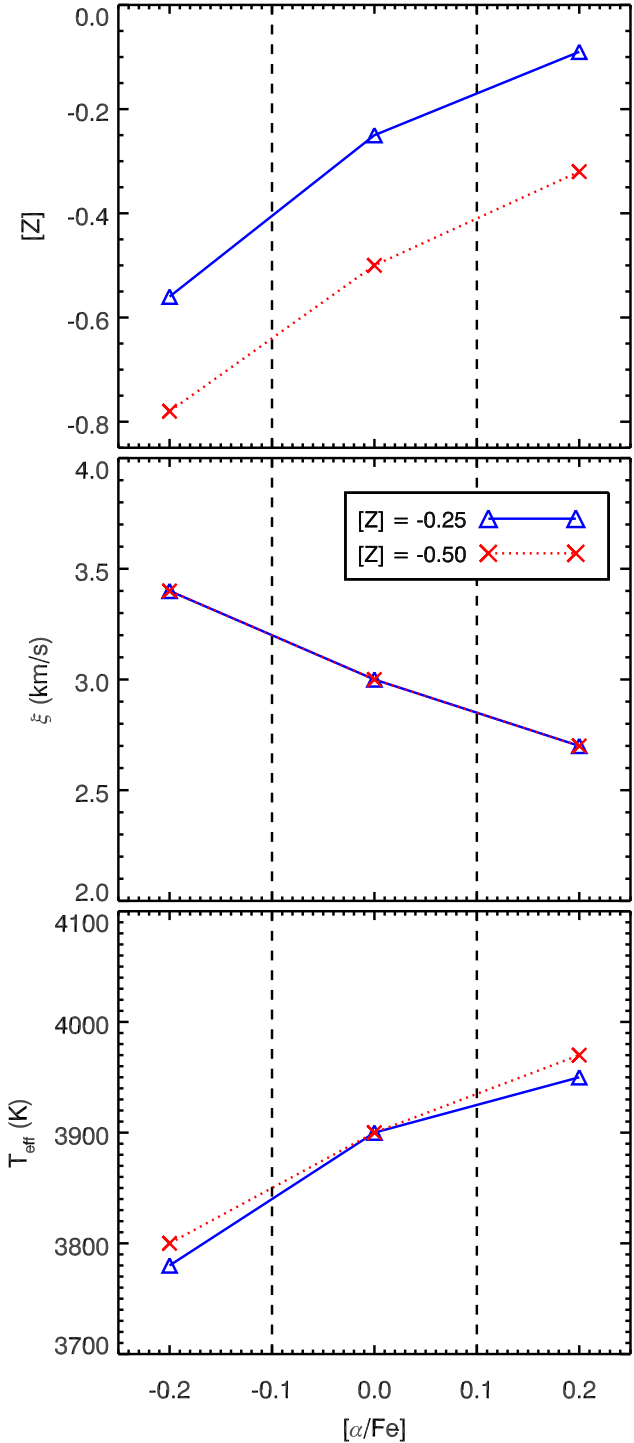}
\caption{The impact on the derived physical properties of a star with non-Solar \afe\ ratios when its spectrum is analysed with a grid of models with Solar \afe. The panels show the inferred metallicity ({\it top}), microturbulence ({\it middle}) and temperature ({\it bottom}) as a function of the input star's \afe\ ratio, for two global metallicities. See text for details.}
\label{fig:afe}
\end{figure}

\subsection{Departures from Solar-scaled $\alpha$/Fe } \label{sec:afe}

Throughout this study we have consistently assumed a value of [$\alpha$/Fe]=0.0. Our literature study of analyses of young stars (Tables \ref{tab:lmc} and \ref{tab:smc}) supports this assumption, appearing to show that [$\alpha$/Fe] for the two Magallanic Clouds is Solar to within $\pm$0.1dex, with the results of the FLAMES survey of massive stars indicating that perhaps [Si/Fe] and [Mg/Fe] were depleted by 0.1-0.2dex with respect to Solar. However, a comprehensive study of BA supergiants in the SMC by \citet{Schiller_thesis} concluded that \afe\ was {\it enhanced} by 0.1dex with respect to the Solar abundances of \citet{GAS07}. 

Here, we now explore the effect that a small departure from Solar [$\alpha$/Fe] would have on our results if Solar \afe\ were incorrectly assumed. To do this, we constructed a model spectra with parameters [Z]=-0.5, \teff=3900K, $\xi$=3\kms, \logg=0.0, and with \afe=$\pm$0.2. We then analysed these spectra in the same way as in the rest of this work using the models with Solar \afe. 

In \fig{fig:afe} we plot the degeneracy between \afe\ and the model parameters [Z], $\xi$ and \teff\ (there was no detectable degeneracy with \logg). In the bottom panel we see there is a small trend between \teff\ and \afe. This can be explained by the fact that the relative strengths of species of lines of different excitation potentials are a diagnostic of temperature (see Sect.\ \ref{sec:teff}). By altering the relative strengths of, for example, \fei\ and \sii\ lines, we may mimic the spectrum of a star with a slightly different temperature with Solar-scaled abundances. This effect however is small -- increasing \afe\ by 0.2dex results in a change of inferred \teff\ of only $\sim$50K.

It is a similar situation for microturbulence. This parameter is sensitive to the relative strengths of the strong and weak lines. In principle this could be estimated from the lines of just one element, for example \fei\ as discussed in Sect.\ \ref{sec:xi}. In practise, our holistic \chisq-minimization approach considers all lines together to increase the precision on each parameter, which results in the degeneracy seen in the middle panel of \fig{fig:afe}. An increase in \afe\ of 0.2dex would result in $\xi$ being underestimated by $\sim$0.3\kms.   

The effect on overall metallicity is predictable -- altering the abundances of the $\alpha$ elements, which constitute six of the eight diagnostic lines, results in an inferred metallicity which is of order that by which the $\alpha$ elements are altered. The small changes in \teff\ and $\xi$ are required to alter the strengths of the \fei\ lines relative to those of the \sii\ and \tii\ lines in order to provide a better fit to the input spectrum.

The dashed vertical lines in \fig{fig:afe} illustrate the possible deviations from Solar \afe\ indicated by our literature search ($\pm$0.1dex). If \afe\ were to be enhanced by 0.1dex, as suggested by \citet{Schiller_thesis}, we would expect to see overall metallicities enhanced by a similar amount with respect to other studies, whilst we might also see average \teff\ and $\xi$ values which were slightly higher {\it if} the average values of these quantities were independent of metallicity, which may or may not be the case. Our results for the SMC do show slightly higher abundances than other works by around 0.1dex, while the average $\xi$ value does seem to be lower by a few tenths of \kms. This is circumstantial evidence, albeit rather weak, for a small \afe\ enhancement in the SMC. 

In the near future we plan to undertake a more thorough analysis of \afe\ ratios in these galaxies. This will be possible once we have implemented non-LTE corrections to the two \mgi\ lines in our spectral window (Bergemann et al., in press), {with the increased number of diagnostic lines and a third $\alpha$-element enabling us to separate \afe\ from [Fe/H] in our abundance analysis.}




\begin{table*}[htdp]
\caption{Average LMC abundances}
\begin{center}
\begin{tabular}{lccccl}
Target(s) & [O] & [Si] & [Mg] & [Fe] & Ref. \\
\hline
{\it B dwarfs} \\
N11      & -0.34$\pm$0.08 & -0.38$\pm$0.07 & -0.47$\pm$0.09 & -0.22$\pm$0.10$\dagger$&\citet{Dufton06}\\
NGC~2004 & -0.27$\pm$0.18 & -0.36$\pm$0.03 & -0.45$\pm$0.10 & -0.29$\pm$0.12$\dagger$&\citet{Trundle07}
\smallskip \\
{\it Cool Supergiants} \\ 
F supergiants$\dagger$ & -0.22$\pm$0.08 & +0.08$\pm$0.10 & -0.32$\pm$0.08 & -0.22$\pm$0.06 & \citet{Hill95} \\
" " (same sample) & -0.16$\pm$0.08 & - & - & -0.34$\pm$0.15$\dagger$ & \citet{Andrievsky01} \\
Cepheids$\dagger$ & -0.01 & -0.12 & - & -0.14$\pm$0.12 & \citet{Luck98} \\ 
Cepheids$\dagger$ & - & - & - & -0.27$\pm$0.12 & \citet{Romaniello08} \\
\smallskip \\
\hline
\it Solar & 8.66 & 7.51 & 7.53 & 7.45 & \citet{GAS07} \\
\hline
\end{tabular}
\end{center}
$\dagger$Study did not account for non-LTE effects on diagnostic lines.
\label{tab:lmc}
\end{table*}%

\begin{table*}[htdp]
\caption{Average SMC abundances}
\begin{center}
\begin{tabular}{lccccl}
Target(s) & [O] & [Si] & [Mg] & [Fe] & Ref. \\
\hline
{\it B dwarfs} \\
NGC~330  & -0.78$\pm$0.20 & -0.67$\pm$0.02 & -0.80$\pm$0.10 & - &\citet{Trundle07} \\
" "          & -0.69$\pm$0.13 & -0.91$\pm$0.32 & -0.91$\pm$0.14 & - &\citet{Lennon03} \\
" "          & -0.42$\pm$0.30 & -0.68$\pm$0.40 & -0.69$\pm$0.30 & - &\citet{Korn00} \\
NGC~346  & -0.61$\pm$0.10 & -0.71$\pm$0.05 & -0.76$\pm$0.07 & - &\citet{Trundle07} \\
AV~304      & -0.54$\pm$0.13 & -0.76$\pm$0.16 & -0.79$\pm$0.20 & - &\citet{Hunter05} \smallskip \\
{\it Hot Supergiants} & & & & & \\
field B SGs & -0.54$\pm$0.13 & -0.75$\pm$0.18 & -0.69$\pm$0.14 & - & \citet{Trundle05} \\
field A SGs$\dagger$ & -0.52$\pm$0.06 & -0.54$\pm$0.18 & -0.78$\pm$0.10 & -0.80$\pm$0.15 & \citet{Venn95,Venn99} \smallskip\\
{\it Cool Supergiants} \\ 
K SGs (NGC~330)$\dagger$ & -0.9 & -0.8 & -0.9 & -0.8 & \citet{Hill99} \\
field K SGs$\dagger$ & -0.7 & -0.8 & -0.5 & -0.6 & \citet{Hill97a,Hill97b} \\
Cepheids$\dagger$ & -0.6 & -0.5 & -0.6 & -0.6 & \citet{Luck98} 
\smallskip \\
\hline
\it Solar & 8.66 & 7.51 & 7.53 & 7.45 & \citet{GAS07} \\
\hline
\end{tabular}
\end{center}
$\dagger$Study did not account for non-LTE effects on diagnostic lines.
\label{tab:smc}
\end{table*}%

%
%
%
%

\section{Summary \& conclusions} \label{sec:conc}
We have presented a metallicity study of the Large and Small Magellanic Clouds (LMC and SMC respectively) using VLT/XSHOOTER near-IR spectroscopy of samples of Red Supergiants (RSGs). Such stars are young and their abundances of Mg, Ti, Si and Fe accurately reflect those in the gas phase. We concentrate our analysis on a narrow window in the $J$-band, where molecular absorption is weak, and where we have shown previously that accurate stellar abundances may be obtained. Our results can be summarised as follows:

\begin{itemize}
\item Our analysis of the two samples of stars reveal metal abundances of elements heavier than He relative to Solar of [Z]=-0.37$\pm$0.14 for the LMC and [Z]=-0.53$\pm$0.16 for the SMC. Both results are consistent with other studies of young massive stars in the literature, though the SMC result is at the high end of that found by other comparable works ($\sim$0.5-0.8dex below Solar). 
\item We find best-fitting temperatures which are consistent with those from fits to the optical-infrared spectral energy distribution, though there is a small systematic offset of marginal statistical significance for the LMC stars, 160$\pm$110\,K. 
\item The average microturbulent velocities $\xi$ in the LMC (3.3$\pm$0.5\kms) are consistent with those found by ourselves in Galactic stars. The average for the SMC stars is slightly lower at 2.8$\pm$0.3\kms. Though a low-significance result, we offer two possible explanations: firstly, it may be indicative of the physics of convection at low metallicity; or secondly, it could be a systematic effect caused by the SMC RSGs having slightly super-Solar \afe\ ratios.   
\end{itemize}

In the near future we will explore the effect of non-Solar \afe\ ratios on our results by incorporating non-LTE corrections to the $J$-band \mgi\ lines into our analysis. The increased number of diagnostic lines {and atomic species} will then enable the ratio of $\alpha$-elements to be considered separately to the Fe lines. We will also extend our metallicity baseline to lower [Z] systems such as WLM, to {further investigate any potential trend of $\xi$ with decreasing [Z].}

\acknowledgments Acknowledgments: We are very grateful to the anonymous referee who helped us improve our manuscript. BD was funded in part by a fellowship from
the Royal Astronomical Society. RPK and JZG were supported in part by the National
Science Foundation under grant AST-1108906 (PI: R.-P. Kudritzki). Moreover, RPK
acknowledges support by the Alexander-von-Humboldt Foundation and the
hospitality of the Max-Planck-Institute for Astrophysics in Garching
and the University Observatory Munich, where part of this work was
carried out. For our analysis we used the software package IDL, the The IDL Astronomy User's Library at GSFC, and the Coyote graphics library.

\bibliography{../biblio}

\end{document}